\documentclass[12pt,preprint2]{aastex}

\begin{document}

  \title{Characterizing Charge Diffusion in CCDs with X-rays}
  \author{Steven A. Rodney}
  \affil{Institute for Astronomy, University of Hawaii \\
  	Honolulu, HI 96822}
  \email{rodney@ifa.hawaii.edu}

  \author{John L. Tonry}
  \affil{Institute for Astronomy, University of Hawaii \\
    Honolulu, HI 96822}
  \email{jt@ifa.hawaii.edu}

  \begin{abstract}
    We demonstrate the effectiveness of two techniques for using
    x-rays to evaluate the amount of charge diffusion in charge
    coupled devices (CCDs). We quantify the degree of charge diffusion
    with two parameters: $\sigma_d$, the standard deviation for a
    Gaussian diffusion model, and Q, a ratio of the point
    spread function (PSF) peak to its wings. $\sigma_d$ and Q are
    determined by fitting a model to a pixel energy histogram, and by
    summing the PSF of all x-ray events, respectively. Using seven test
    devices, we investigate the precision of these two techniques and
    demonstrate that they produce compatible results. The histogram
    fitting method is sensitive to the structure of the electric field
    within these devices, in addition to the inherent charge diffusion
    properties. The Q ratio is a very simple parameter to measure and
    provides an easily accessible method for quickly evaluating a
    CCD's diffusion length.  
  \end{abstract}


  \section{Introduction} \label{intro}
  
  The use of Charge Coupled Devices (CCDs) as light collectors for
  photometry and spectroscopy is ubiquitous in modern observational
  astronomy. To understand fully the limits and capabilities of these
  devices it is extremely important to have a set of standard methods
  for testing and quantifying their properties. There are a number of
  useful CCD characteristics that are relatively straightforward to
  measure.  For example, carefully refined techniques for testing the
  gain, linearity, and charge transfer efficiency (CTE) have been
  developed \citep{jan2001}.  One important CCD trait for which an
  efficient and practical testing procedure does not exist is charge
  diffusion.

\begin{figure}[tb]
  \centering
  \includegraphics[width=0.4\textwidth,draft=false]{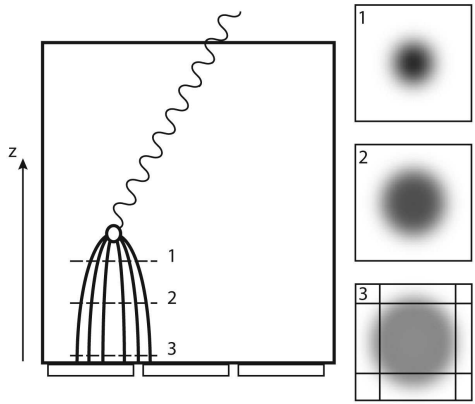}
  \caption{{\small Cartoon representation of an x-ray absorption in a CCD
      and the resulting charge diffusion cloud.  An x-ray enters
      through the backside surface at the top, and interacts at some
      height z above the gates. The initial size of the electron cloud
      is small compared to the size of the pixels. As the electrons are 
      pulled downward at a velocity proportional to the electric field
      strength they random walk horizontally. A cross-section of the
      cloud at any height z in the silicon looks approximately like a
      two dimensional Gaussian distribution of electrons (insets 1, 2,
      and 3). When the cloud reaches the bottom of the device it is
      segmented by the grid of the pixel structure.
      \label{fig:chdiff_cartoon}}}
\end{figure}

  Figure ~\ref{fig:chdiff_cartoon} depicts schematically the process of
  charge creation and collection within a CCD.
  When a photon reaches the surface of a CCD it penetrates through
  some depth of silicon before it is absorbed. At the point of
  interaction this photon liberates one or more charge carriers.  A
  photon with energy less than $\sim$4 eV creates one charge
  carrier. Higher energy x-rays produce a packet of charge in which
  the number of charge carriers is proportional to the energy of the
  photon. These charge carriers are either electrons or holes,
  depending on the CCD design.  Hereafter we will assume that the
  buried channel is in an n-type semiconductor, so the charge carriers
  are electrons. The electron thermal velocities cause them to wander
  away from their original location
  into neighboring pixels.  This spreading effect is what is termed
  charge diffusion.  The electrons produced by photon absorption are
  eventually converted into a digital image, so charge diffusion
  places a fundamental physical limit on the resolution in the final
  image produced by a CCD. 
  
  \citet{kra1995} characterized charge diffusion in the process of
  evaluating quantum efficiency in x-ray CCDs, but their method
  involved a synchrotron source to produce a continuum of x-ray
  energies. \citet{jan2001} also describes several techniques, such as
  measuring the modulation transfer function by projecting a
  sinusoidal light pattern across the CCD. In general, these methods
  can be effective, but they require either expensive specialized
  equipment or a significant investment of time. Such procedures are
  better suited to CCD manufacturers or engineering labs. 
  Despite the demonstrated importance of understanding charge
  diffusion effects when designing astronomical instruments, there is
  currently no method for effectively evaluating charge diffusion that
  is accessible to the astronomical community without specialized
  equipment.
	 
  The techniques presented here rely on the use of soft x-rays from a
  mild radioactive source as a means of producing discrete, single
  photon events.  These events are approximately point sources,
  since the size of  the initial charge cloud is generally very small
  compared to the size of the pixels \citep{tsu1999}. 
  Of course, the experimenter has no control over where the x-rays
  land on the CCD surface or at what depth in the silicon each photon
  produces charge, so the x-ray events are randomly distributed
  throughout the  volume of the CCD.  Thus, we can gain little
  information by fitting a model to an individual event. An x-ray
  interacting near the surface of a low-diffusion device will spread
  charge over a large area, much like an x-ray interacting deeper down
  but in a device with higher charge diffusion.    We can,
  however, break this degeneracy by modeling the entire ensemble of
  x-ray events and including the absorption probability as a function
  of depth (i.e. absorption probability increases exponentially with
  depth, so there will be far more interactions deep in the silicon
  than near the surface). For the two methods described here, we
  compare the histogram of pixel energies or the stacked and summed
  x-ray PSF to models that incorporate all of the available
  information about the charge collection process.

  In the following section we describe the experimental setup and data
  collection procedures, as well as the formula for the calculation of
  the Q ratio. \S \ref{sec:xhist} details the creation
  and interpretation of K-$\alpha$ pixel histograms, while \S
  \ref{sec:model} explains the theoretical model used to fit those
  histograms.  We present results for the two methods in \S
  \ref{sec:results} and final conclusions in \S
  \ref{sec:conclusions}. 

  \section{Experimental Setup and Procedures} \label{setup}

  Seven devices, labeled A-G, were used for the diffusion measurements. 
  The first three are CCID-20 chips, chosen
  primarily because of their disposability - these CCDs were not
  likely to be used in a science-grade instrument, so damages incurred
  while testing them would not have been disastrous.  The other four
  are CCID-45 Orthogonal Transfer Array (OTA) devices \citep{ton1997}.
  Each OTA has 64 individually addressable cells of approximately
  500x500 pixels each.  Table \ref{tab:ccdlist} lists the relevant
  information for  all of the devices used.   This set of seven CCDs
  provides a useful variety of physical parameters that affect the
  measured charge diffusion, such as silicon thickness and pixel size.
  The OTAs have the unique additional feature of allowing the voltage
  level of the substrate to be adjusted during testing, which provides
  another way to modify the diffusion properties while holding all
  other parameters constant.  

\begin{table*}
  \caption{Devices Tested}
  \smallskip
  \begin{center}
    {\small
      \begin{tabular*}{\textwidth}{c c c c c c l}
	\tableline
	\tableline
	Device 
	& Model
	& Thick.
	& Pix. Size
	& $h_{bf}$\tablenotemark{*}
	& Resistivity
	& Notes \\
	
	(CCD-)
	& (CCID-)
	& $(\mu m)$
	& $(\mu m)$
	& $(\mu m)$
	& $(k\Omega-cm)$
	& \\
	
	\noalign{\smallskip}
	\tableline
	\noalign{\smallskip}
	
	A & 20 & 20 & 15 & 4 & 0.3 & bright defects; useable area $\sim$85\%\\
	B & 20 & 45 & 15 & 6 & 5 & bad parallel CTE along one side; \\
	&  & & & & & useable area $\sim$90\%\\
	C & 20 & 45 & 15 & 8 & 5 & tree-ring pattern fab. error \\
	D & 45 & 45 & 10 & 10 & 5 & variable substrate voltage\\
	E & 45 & 45 & 12 & 10 & 5 & variable substrate voltage \\
	F & 45 & 45 & 12 & 12 & 14 & variable substrate voltage \\
	G & 45 & 75 & 12 & 14 & 14 & variable substrate voltage \\
	
	\noalign{\smallskip}
	\tableline
	\noalign{\smallskip} 
	
	\tablenotetext{*}{From histogram fits. See section \ref{sec:model}}
	
      \end{tabular*}
    }
  \end{center}
  \label{tab:ccdlist}
\end{table*}

\begin{figure}[tb]
  \centering
  \includegraphics[width=0.4\textwidth,draft=false]{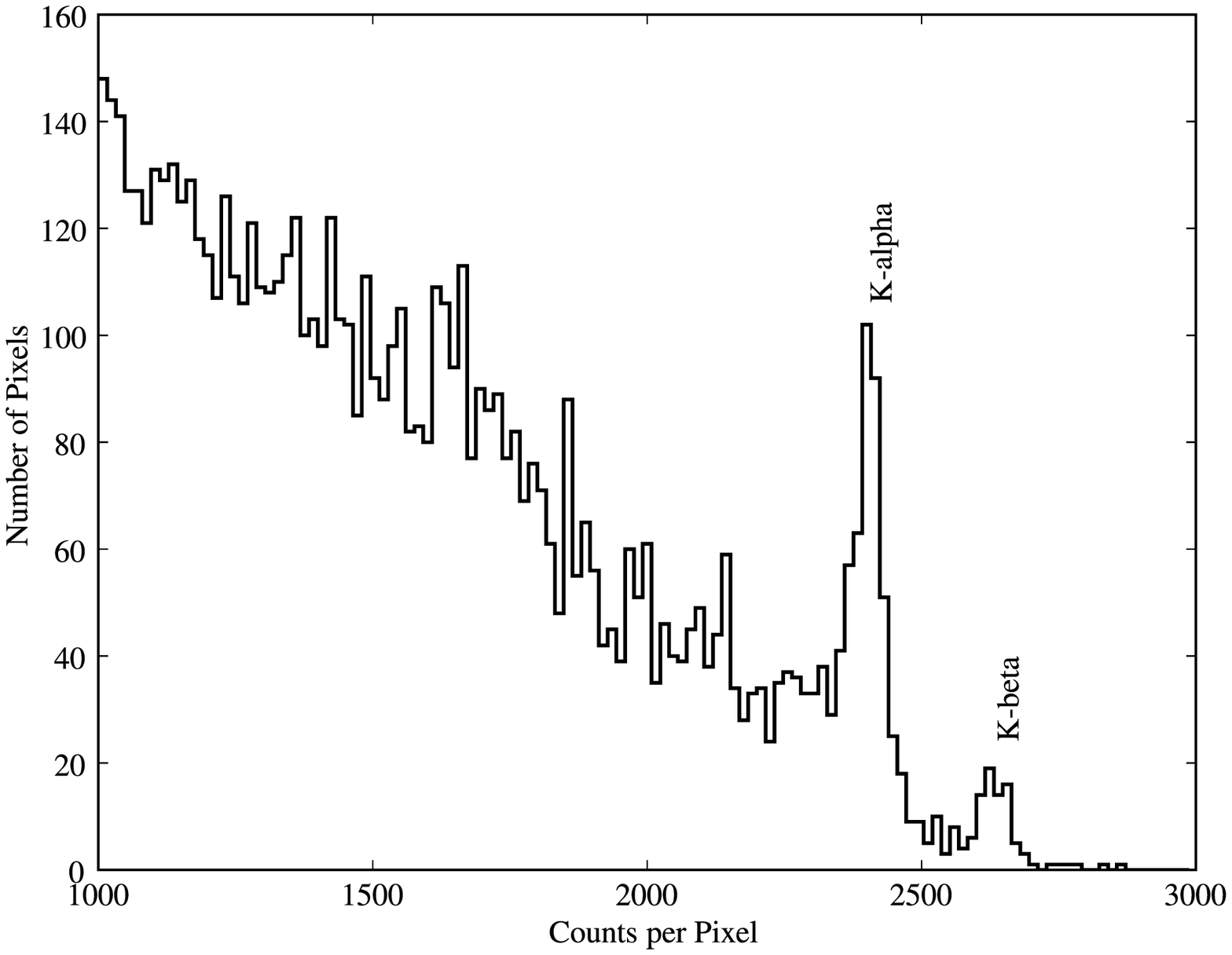}
  \caption{\small
      The complete pixel value histogram for a typical x-ray
      image. The Mn K-$\alpha$ and K-$\beta$ peaks are known to
      produce 1620 and 1778 electrons, respectively, so the gain of
      this device can be easily determined by measuring the peak
      location in ADU and dividing the appropriate electron number by
      this value.  
      \label{fig:rawhist}}
\end{figure}

  To provide soft x-ray illumination, we use a 1 mCi Fe55 x-ray
  source. The Fe55 isotope emits x-rays when the nucleus captures an
  inner electron and becomes manganese. The most important x-ray
  emissions are the Mn K-$\alpha$ event at 5.899 keV and the K-$\beta$
  line at 6.490 keV. K-$\alpha$ and K-$\beta$ x-rays interact in the
  silicon and produce a characteristic charge cloud equivalent to 1620
  and 1778 electrons, respectively. Each of the four devices is
  mounted in a dewar that has a beryllium window in its faceplate,
  opaque to optical and infrared light but transparent to
  x-rays. By placing the Fe55 x-ray source directly against this
  window while the CCD is integrating, we get a dark image with
  hundreds of isolated single x-ray events. 

  Figure \ref{fig:rawhist} shows a simple histogram of all pixel
  values from an x-ray image of CCD-A. This  histogram includes
  every pixel in the image (above a threshold of 1000 ADU),
  regardless of whether it is identified as a K-$\alpha$ pixel or not.
  Such histograms are useful for accurately measuring the gain of a
  device, but to measure charge  diffusion we really want to limit
  ourselves to those pixels in the immediate vicinity of an x-ray
  event. To produce the required x-ray pixel histogram, an x-ray image
  is scanned with a point source locating algorithm, which locates all
  pixels that are adjacent to a K-$\alpha$ x-ray event.
  This algorithm first creates a raw histogram, such as the one in
  Figure \ref{fig:rawhist}, and locates the K-$\alpha$ peak to determine
  $N_{k\alpha}$, the number of ADUs that corresponds to a complete
  K-$\alpha$ charge cloud  (i.e. 1620 electrons times the gain of the
  device). We then examine all pixels that have a signal more than
  seven times the RMS noise of the background, along with their eight
  immediately adjacent pixels.  For each patch of nine pixels, we
  subtract a constant background level and calculate the sum of nine
  pixel values, $N_{patch}$.  If this sum is close to $N_{k\alpha}$,
  meaning  $$\vert N_{k\alpha} - N_{patch} \vert <  \frac{(N_{k\alpha}
  - N_{k\beta})}{2} $$ then all nine pixels are tagged as 'K-$\alpha$
  pixels' and are included in the calculation of Q and in the
  x-ray pixel histogram. 

  To measure Q, the 3x3 arrays of K-$\alpha$ pixels, 
  centered on the brightest pixel, are
  stacked and summed to produce a composite x-ray PSF.  Q is
  then defined as the normalized ratio of the sum of the eight outer
  pixel values to the peak pixel value in the composite PSF: 

  \begin{equation}\label{eq:qdiff}
    Q = \frac{1}{8}\frac{\sum{(eight~outer~pixels)}}{(central~pixel)}
  \end{equation}

  \noindent Thus, for a device with zero charge diffusion $-$ so that
  all 1620 electrons produced by a K-$\alpha$ x-ray  are collected in
  a single pixel every time $-$ the Q ratio is zero.  For
  infinite charge diffusion $-$ so that each of the nine pixels
  receives an equal fraction of the total charge $-$ Q is
  unity.

  \section{The X-ray Pixel Histograms} \label{sec:xhist}

\begin{figure}[!tb]
  \centering
  \includegraphics[width=0.36\textwidth,draft=false]{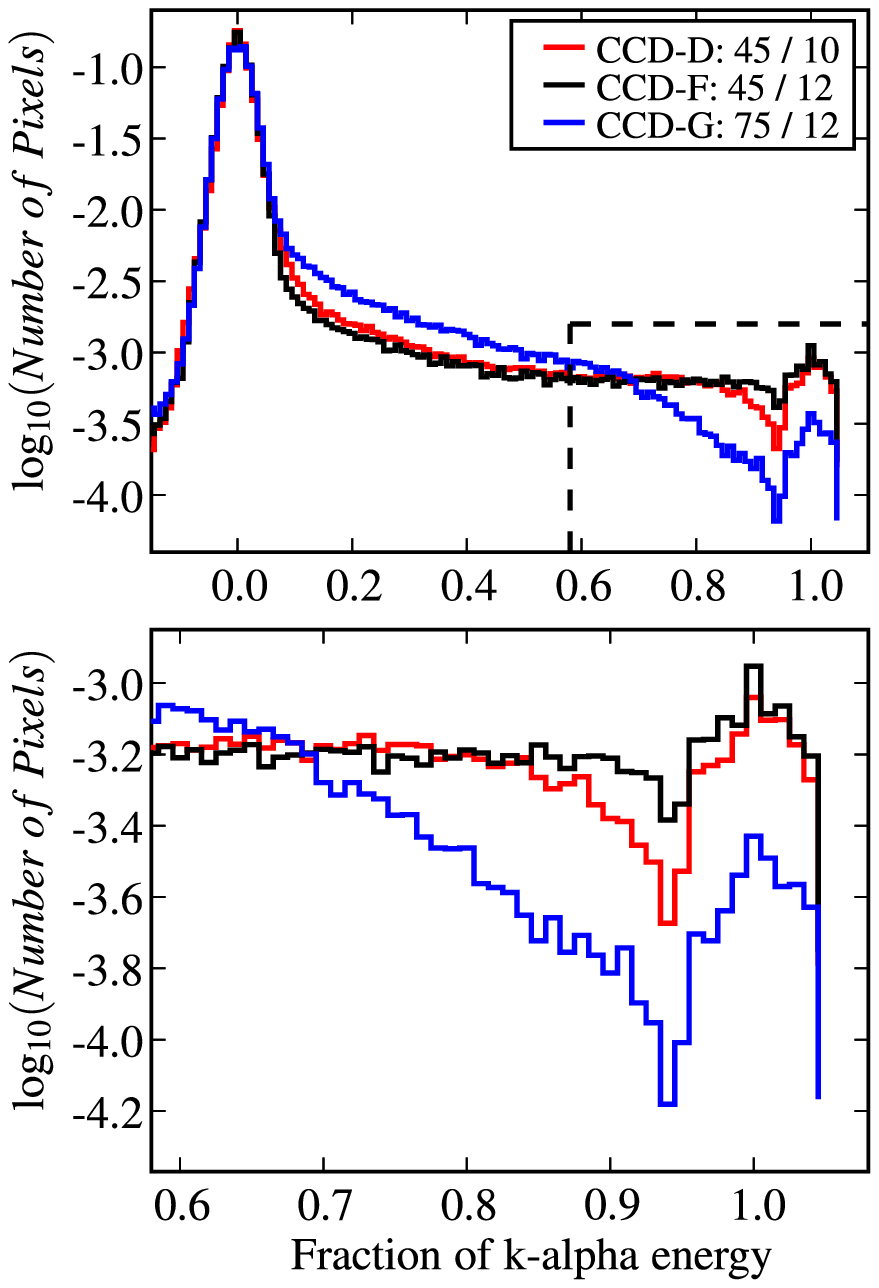}
  \caption{{\small
      Logarithmic x-ray histograms from CCD-D, F and G in red, black
      and blue, respectively.  All three histograms were created with
      the CCDs set to the same substrate voltage of -10 V. The legend
      in the upper panel gives the thickness and pixel size in $\mu m$
      for each device.  CCD-D and F have the same thickness, but
      different  pixel sizes. CCD-D, with smaller pixels,
      shows a dearth of events with highly concentrated charge
      collection. CCD-F and G have the same pixel size, but different
      thicknesses. The thicker device, CCD-G, has substantially fewer
      x-ray events with more than 70\% of the energy collected by a
      single pixel, relative to the thinner CCD-F. 
    \label{fig:3hist}}}
\end{figure}

  After computing Q, the algorithm then collects the complete set
  of all K-$\alpha$ pixels (meaning all nine pixels from every patch
  centered on a K-$\alpha$ event) and bins them into a histogram
  indicating what  fraction 
  of the x-ray's energy was deposited in each pixel. To facilitate
  comparison with other devices and models, these x-ray pixel
  histograms are normalized such that the integral over all histogram
  bins is unity. These histograms are dominated by the background pixels,
  clustered around zero energy, so we display them on a log scale to
  highlight the subtle features near unity, which are strongly
  affected by the amount of charge diffusion.  Figure
  \ref{fig:3hist} shows three x-ray
  histograms, demonstrating how physical properties of the CCD alter
  the histogram appearance. 

  The prominent features of these x-ray histograms can be explained in
  terms of the physical process of charge diffusion and collection.
  The strong peak around an energy value of zero is of course due to
  the pixels included in each 3x3 patch that have very little or no
  contribution from the nearby K-$\alpha$ electrons.  This peak
  extends into negative energy values because of noise fluctuations
  that leave some pixels below the median background level. To the
  right of this peak, as should be expected, the slope of the
  histogram is mildly negative through the intermediate energy bins.
  The first feature of interest to the determination of charge
  diffusion is the slope discontinuity, which is particularly 
  prominent in CCD-G as shown in Figure \ref{fig:3hist}.\footnote{We
  emphasize here that these  
  histograms are plotted 
  on a logarithmic scale, so these are in fact very subtle slope
  discontinuities.}   This slope discontinuity indicates a sudden
  decrease in the number of pixels containing a large fraction of the
  K-$\alpha$ electrons. The energy at which this critical point occurs
  marks the energy remaining in the central pixel when an x-ray lands
  at the center of a pixel and the surface of a CCD.
  In any CCD, there are essentially two ways for an
  x-ray event to deposit a large fraction of its charge in a single
  pixel: it can land very near the center of the pixel's square
  cross-section (allowing it lots of room to spread out within a
  single pixel), or it can interact with the silicon at a vertical
  height close to the gates (giving it very little time to spread out
  to other pixels).  In CCDs with small pixels (such as
  CCD-D in Fig. \ref{fig:3hist}) the first avenue is less effective,
  so the discontinuity point is translated to the left relative to
  large-pixel devices.  In thick devices, the second effect is
  suppressed because the x-rays must travel through a greater volume
  of silicon to reach the requisite depths for containing the charge
  cloud in one pixel.  Thus, thicker devices (such as  CCD-G in
  Fig. \ref{fig:3hist}), have fewer single pixels holding most of
  the K-$\alpha$ electrons.  

  The final peak of these x-ray histograms is populated by
  single-pixel x-ray events.  These are events in which the entire
  charge cloud is collected in one pixel.  The presence of this peak
  in all of our x-ray histograms indicates that there is a 
  sizeable probability 
  for an x-ray penetrating deep into the CCD to deposit all
  of its energy into one pixel, rather than to spread a small fraction
  of it to neighboring pixels.  As explained in \S \ref{sec:model},
  this trait is amplified by the presence of pixel barrier fields,
  which effectively focus charge clouds into pixel wells.

  \section{The Diffusion Model} \label{sec:model}

  To construct a model for comparison with the histograms described
  above, we consider an x-ray incident on the backside of a CCD that
  is absorbed at some height z above the frontside gate, as shown in
  Figure \ref{fig:chdiff_cartoon}. The potential difference across the
  device from the backside to the buried channel produces an electric
  field that draws the electrons into the potential well which defines
  each pixel. If these were free electrons in empty space, they would
  be accelerated by this electric field. Inside a silicon lattice,
  however, these charges are impeded by collisions with lattice atoms
  and it can be shown that the drift velocity is proportional to the
  electric field at that height: 
  $v_z \propto E(z)$. \citep[p.478]{jan2001}  As the electrons are
  gradually pulled downward, they also execute a random walk
  horizontally, so that when they are finally collected in the gate
  structure the result is a substantially more extended profile than
  the original charge cloud. This final charge distribution is
  generally taken to be a Gaussian \citep{hop1987,jan2001}, although
  some calculations have shown that it should be more centrally peaked
  \citep{pav1999,mcc1995}. For simplicity, we restrict ourselves to the
  classic Gaussian charge distribution in all models.

  We can take the RMS horizontal displacement to be proportional to
  the square root of the time it takes to reach the frontside gate at
  z=0: 

  \begin{equation} \label{eq:sig_E}
    \begin{array}{ l @{\propto} l }
      \sigma ~ & ~ t^{1/2} \\
             & ~ \left({\frac{z}{v_z}}\right)^{1/2} \\
             & ~ \left({\frac{z}{E(z)}}\right)^{1/2} \\			 	
    \end{array} 
  \end{equation}

  \noindent In the simplest case, we can take the electric field to be
  constant with depth, so that a device with thickness d and potential
  difference V from frontside to backside has $E(z) = V/d$ and we can
  now write  

  \begin{equation}  \label{eq:sig}
    \sigma(z) = \sigma_d \left(z/d\right)^{1/2} 
  \end{equation}

  \noindent where $\sigma_d$ is the standard deviation of the Gaussian
  diffusion cloud that would result from an x-ray interacting at the
  surface, where z = d.  Now for a given value of $\sigma_d$ and a
  K-$\alpha$ x-ray event at a known height z we can create a final
  Gaussian distribution of charge.  If we also choose an (x,y)
  position for the center of the cloud relative to the pixel boundary,
  we can integrate that final charge distribution over a 3x3 array of
  CCD pixel boundaries. To create a histogram from these pixel values, we
  want to repeat this process for a large number of x-ray events
  spread over an appropriate range of (x,y,z) values. For this, we
  need to know what fraction of x-ray events is likely to occur at any
  given depth in the silicon. K-$\alpha$ photons in silicon have an
  e-folding absorption depth of 27.8 $\mu m$, so in a device of
  thickness d $\mu m$ the fraction of absorption events occuring
  between height $z_1$ and $z_2$ from the bottom is: 

  \begin{equation} \label{eq:frac}
    f = e^{\frac{-(d-z_1)}{27.8 \mu m}} 
    - e^{\frac{-(d-z_2)}{27.8 \mu m}}    
  \end{equation}
 
  The last physical effect that we must consider is the presence of
  barrier fields produced by the gate structure on the frontside of
  each device.  The division of any CCD into pixels necessarily
  requires potential barriers which delineate the individual pixel
  wells. These barriers introduce a transverse electric field that
  prevents electrons very deep in the silicon from crossing over into
  neighboring pixels. We simulate this effect by introducing a barrier
  field height parameter, $h_{bf}$, such that once an electron reaches
  a height $z < h_{bf}$ it can no longer diffuse away from its current
  position.   This effectively truncates the diffusion process at a
  height $h_{bf}$ above the buried channel.  To accommodate this, we
  rewrite Eq. \ref{eq:sig} as:

  \begin{equation}  \label{eq:sigbf}
    \sigma(z) = \left\{ 
    \begin{array}{ l l }
      \sigma_d \left( z / \left( d - h_{bf}\right) \right)^{1/2} & z > h_{bf} \\
                                       0 & z \le h_{bf} \\
    \end{array} \right.
  \end{equation}

  \noindent Thus, in our models, a 75 $\mu m$ thick device with a
  barrier field height of 11 $\mu m$ exhibits similar diffusion
  characteristics to a 65 $\mu m$ thick device with a 1 $\mu m$
  barrier field.  The one important difference is that the final 11
  $\mu m$ of the first device will absorb substantially more x-rays
  than the final 1 $\mu m$ of the second device, meaning that there
  will be more single-pixel x-ray events populating the final
  peak of the x-ray histogram. Hence, the effect of increasing
  $h_{bf}$ in our models is to amplify the final histogram peak at the
  expense of the energy bins just below the peak. 

  To produce a model energy histogram, we divide the silicon thickness
  into thin slabs. For each slab, Eq. \ref{eq:sigbf}
  determines the apparent $\sigma$(z) at that depth, and
  Eq. \ref{eq:frac} sets the relative number of events that have that
  $\sigma$(z).  Next, we create a regular grid of (x,y) values arranged
  across one triangular eighth of the surface of a single pixel.  Due
  to the symmetry of a square pixel, this is effectively the same as
  using a grid that extends across the entire pixel surface.  For each
  (x,y) pair, we use the $\sigma(z)$ appropriate to the current slab
  to create a final Gaussian diffusion cloud projected onto the pixel
  grid.  We then integrate over each pixel to determine the total
  x-ray counts detected for each event, and collect these pixel values
  into a normalized histogram in exactly the way that was done for the
  actual data.  The final result is an x-ray histogram model with a
  known $\sigma_d$ that can be directly compared to the real
  histogram.

\begin{figure}[tb]
  \centering
  \includegraphics[width=0.4\textwidth,draft=false]{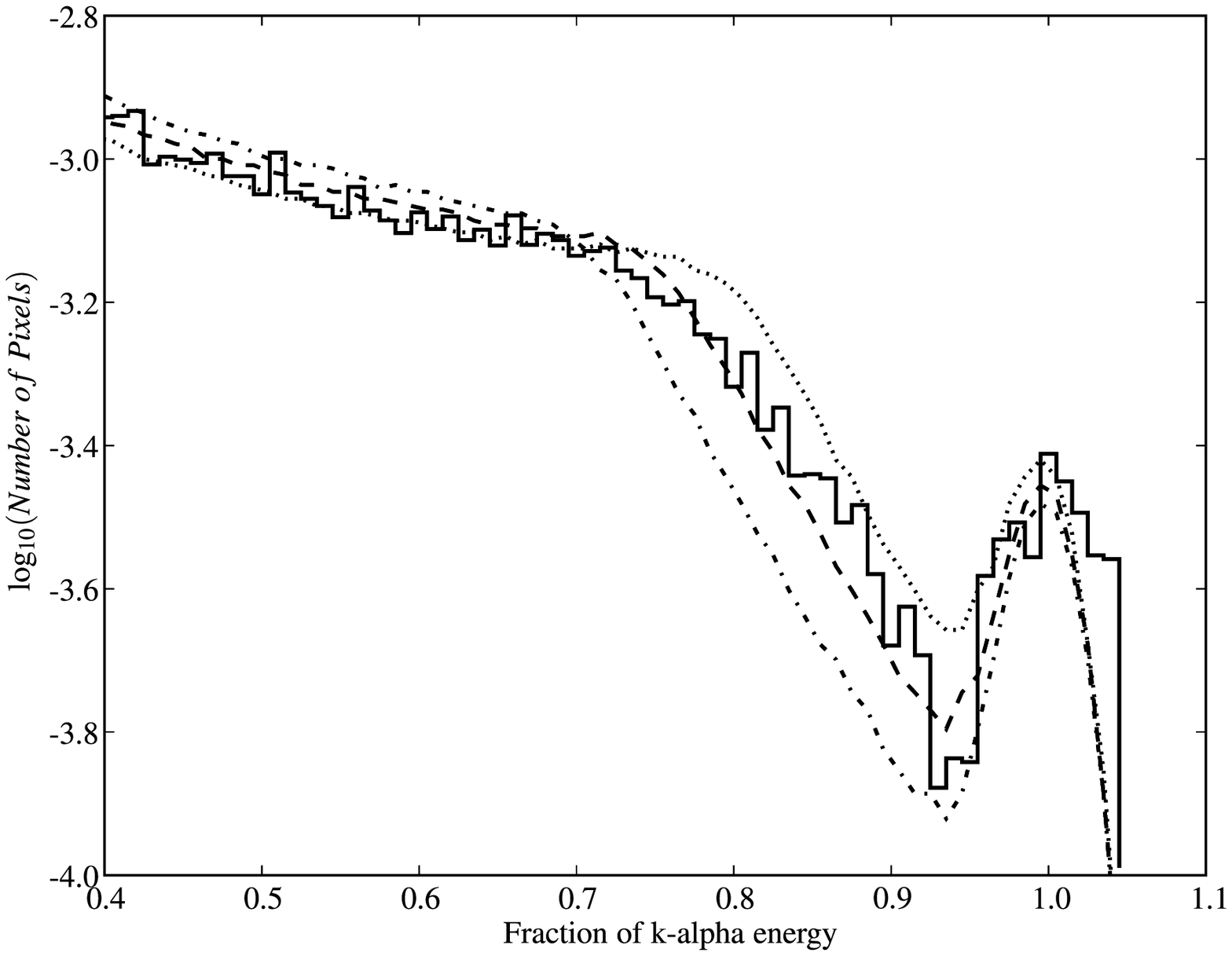}
  \caption{\small
    Demonstration of the precision of our model histogram fitting
    technique.  An actual x-ray pixel value histogram from CCD-G,
    operated with a substrate potential set to -20 V, is
    shown in the solid step-line.   Three model histograms are
    overplotted as dotted, dashed, and dash-dot lines.  These
    models have $\sigma_d$ values of 0.28, 0.30, and 0.32,
    respectively, in units of the pixel width.  For this 12 $\mu m$
    pixel device, these correspond to 3.36, 3.6, and 3.84 $\mu m$. 
    All three models have a barrier field height of $h_{bf} = 14 \mu m$
    \label{fig:precision}}
\end{figure}

  To fit a model to each device we repeat this process for a wide
  range of $\sigma_d$ and $h_{bf}$.  The silicon thickness and
  intrinsic noise are fixed parameters for each device.  From the
  resulting library of model histograms, we choose the curve that most
  closely matches the data histogram and record that model's value of
  $\sigma_d$ as our measured $\sigma_{hist}$.   Figure
  \ref{fig:precision} shows an actual histogram for CCD-G along with
  the three closest matches from our model histogram library.  The
  range in $\sigma_d$ for these three models is 4\% of the pixel
  width, or 0.48 $\mu m$ for this 12 $\mu m$ device.  From these model
  histogram fits we can determine that the precision of
  $\sigma_{hist}$ is $\pm 0.24 \mu m$.  All three models use a barrier
  field height of 14 $\mu m$, which is 19\% of the total 75 $\mu m$
  thickness, and is close to the pixel width.    
  In general, all of the best fitting models for a given
  device agree to within 1 $\mu m$ on the height of the barrier field,
  regardless of the substrate voltage.

  The measured barrier field heights for these seven devices are of
  the order of the size of the pixels.  As a general rule, changing
  the barrier field height in our model has no effect on the
  position of the slope discontinuity point, which is the primary
  indicator of $\sigma_d$.  The height of the final peak and the
  depth of the preceding trough in each histogram are the sole
  indicators of $h_{bf}$.  In devices with very low charge
  diffusion, such as the 20 $\mu m$ thick CCD-A, our determination of
  $h_{bf}$ is less precise, because x-ray events have to be very close
  to the pixel boundaries for the barrier field to have any effect due
  to the inherently small size of the diffusion cloud. These
  low-diffusion devices will also yield somewhat less accurate
  measures of $\sigma_{hist}$.  When there is very little charge
  diffusion the slope discontinuity point that indicates the value of
  $\sigma_{hist}$ is very close to the final peak in the histogram,
  and can therefore be confused by the presence of the preceding
  trough feature. Table \ref{tab:ccdlist} lists the barrier field
  heights measured from histogram fitting in column 5. 

\begin{figure}[tb]
\centering
\includegraphics[width=0.4\textwidth,draft=false]{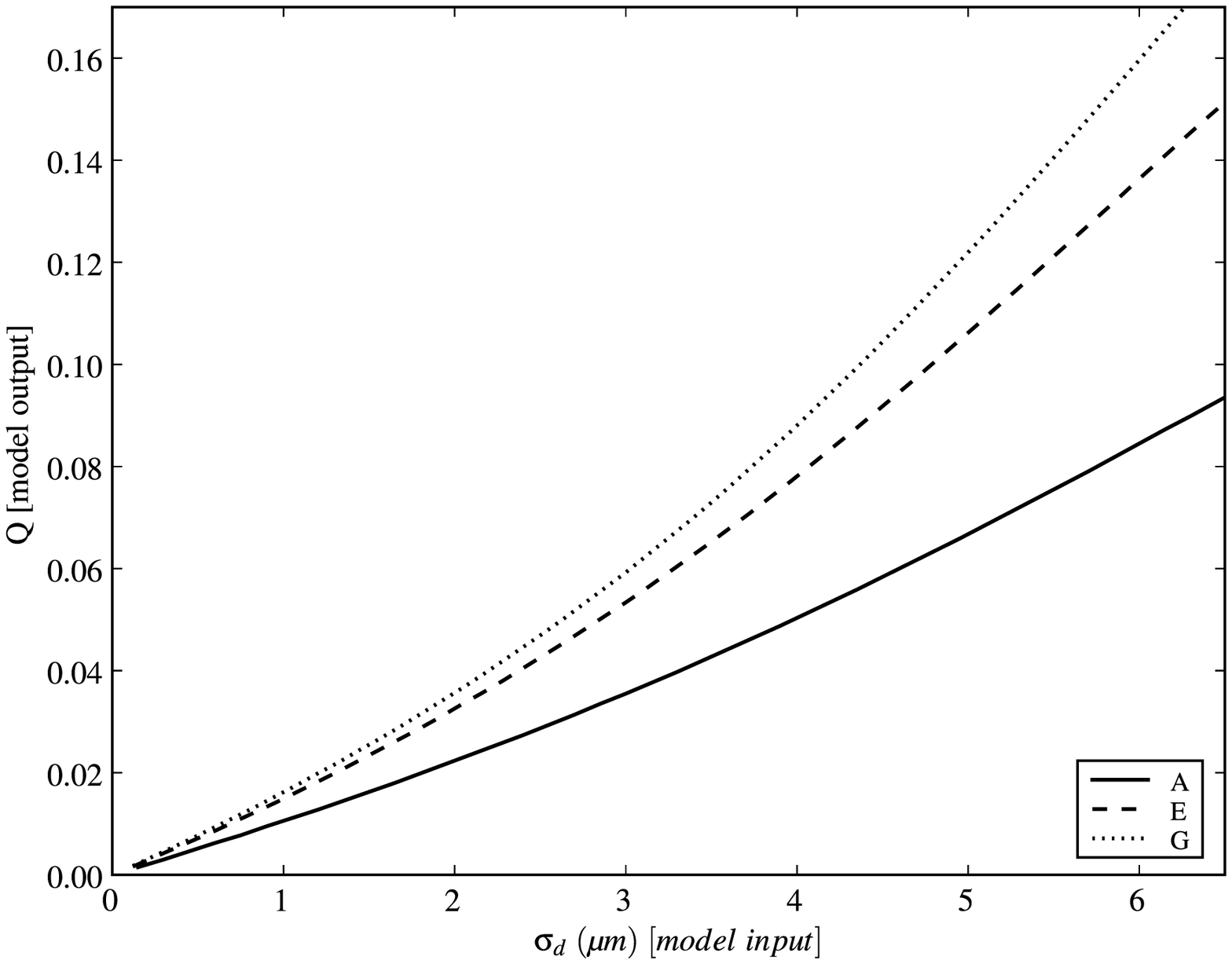}
\caption{\small
      The results of our model recreation of the Q calculation
      process for three of the seven CCDs.  
      For each CCD, the thickness, pixel size and barrier field height
      are set as fixed parameters. Then a series of $\sigma_d$ values
      are input, and our model creates a summed PSF made up of nine
      pixels. This model summed PSF is evaluated to get Q, which is
      then plotted here against the input value of $\sigma_d$.  These
      curves are used to translate our measured values of Q into a
      corresponding $\sigma_Q$ in $\mu m$.  
\label{fig:qsigmod}}
\end{figure}

  For the Q measurements, we simply need to translate our
  measured Q ratio into a physical measurement that can be directly
  compared to the $\sigma_{hist}$ value in microns.  For this purpose
  we employ the same setup to produce a model aggregate PSF, and
  perform the sum and division of the nine pixels to get a value of
  Q. For each device the three physical properties of thickness, pixel
  size, and barrier field height are used as parameters for the model
  to define a curve that shows the input value of $\sigma_d$ plotted
  against the output Q. Figure \ref{fig:qsigmod} shows these curves
  for three of the seven devices tested.    From this plot 
  we can take our unitless measured Q value and translate it directly
  into a value of $\sigma_Q$ in $\mu m$.

  \section{Results} \label{sec:results}

  \begin{figure}[tb]
    \centering
    \includegraphics[width=0.4\textwidth,draft=false]{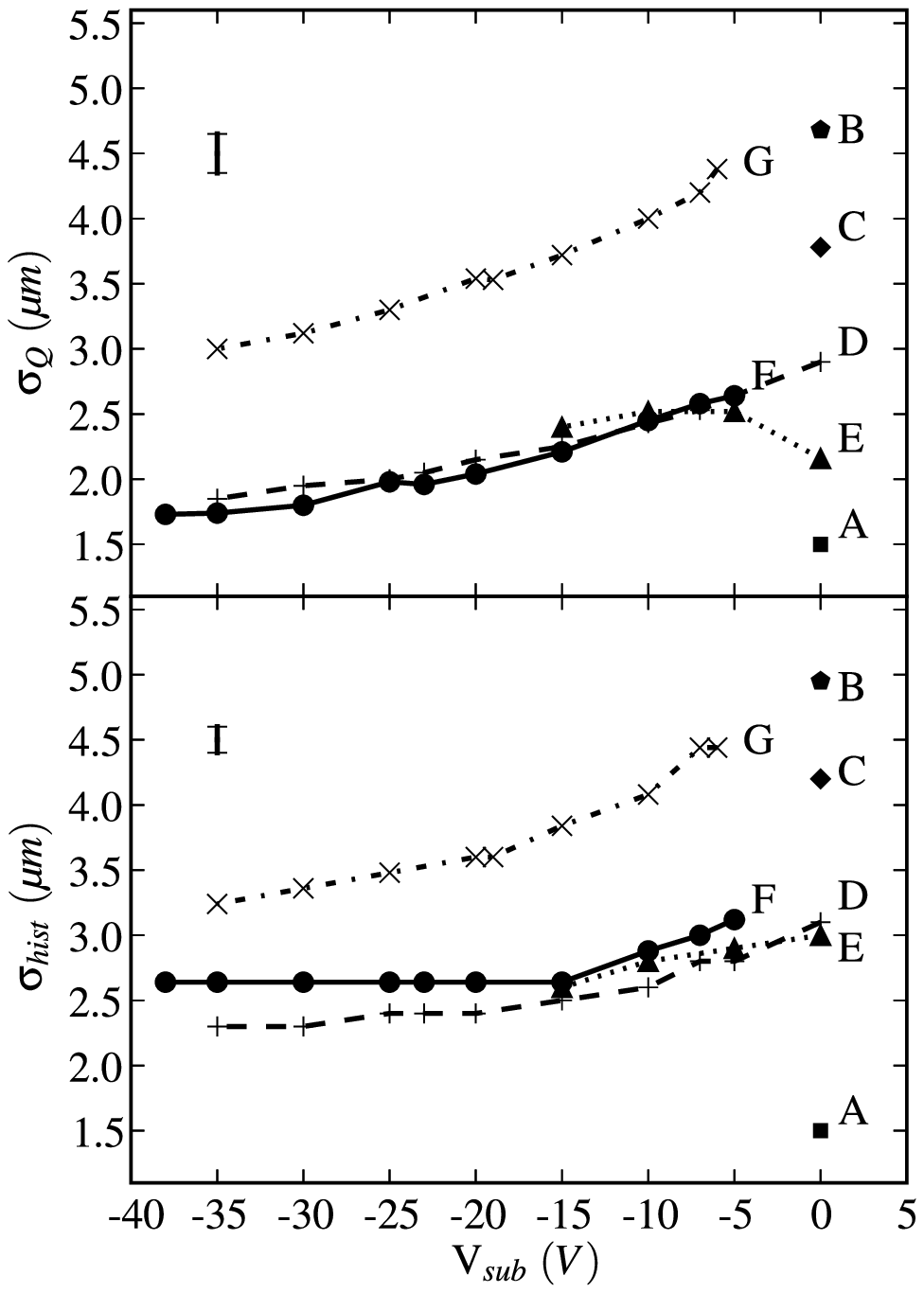}
    \caption{\small
      $\sigma_{Q}$ and $\sigma_{hist}$ plotted as a function of substrate
      voltage.  The non-OTA CCID-20 devices are plotted as single points
      along $V_{sub}$ = 0 V.   A single letter on the right side of the
      figure labels each curve with the corresponding device name.
      Typical error bars are shown in the upper left corners. 
      \label{fig:qsig_vsub}}
  \end{figure}

  Using the procedures described above, we determine a value of
  $\sigma_{hist}$ from histogram fitting, and a value of $\sigma_Q$
  from the summed PSF for each device (and at multiple substrate
  voltages in the case of the OTAs).  These results are shown in
  Figure \ref{fig:qsig_vsub}.  The upper panel of Figure
  \ref{fig:qsig_vsub} plots the $\sigma_{hist}$ values determined from
  histogram fitting against the substrate voltage for all seven
  devices.  The three non-OTA devices (CCD-A, B and C), are plotted as
  single points at $V_{sub}$ = 0.   The lower panel of Figure
  \ref{fig:qsig_vsub} provides the same plots for $\sigma_Q$ as a
  function of substrate voltage.    

  From these plots we can discern a number of important effects.
  First, our two methods both verify that thickness has a strong
  affect on a device's charge diffusion.  The 75 $\mu m$ thick CCD-G
  has almost a factor of two more charge diffusion than the 45 $\mu m$
  OTA devices D, E, and F.
  CCD-A, at a thickness of 20 $\mu m$ has more than a factor 
  of two less diffusion than the 45 $\mu m$ CCD-B and C.  Secondly,
  the distinct histogram shapes resulting from differences in pixel
  size that are shown in Figure \ref{fig:3hist} produce the same
  physical diffusion length when we take into account the pixel size
  in our translation to $\sigma_d$ in $\mu m$. Thus, the curves
  for CCD-D and F are 
  almost identical to within the error bars.  From the four OTAs we
  can evaluate the effects of changing the substrate voltage.  As
  expected, a more negative substrate bias establishes a stronger
  electric field, which results in lower measured
  values of the charge diffusion.  This is true for all devices across
  the whole range of sampled voltages (except for one erratic point on
  CCD-E).  

  \begin{figure}[tb]
    \centering
    \includegraphics[width=0.4\textwidth,draft=false]{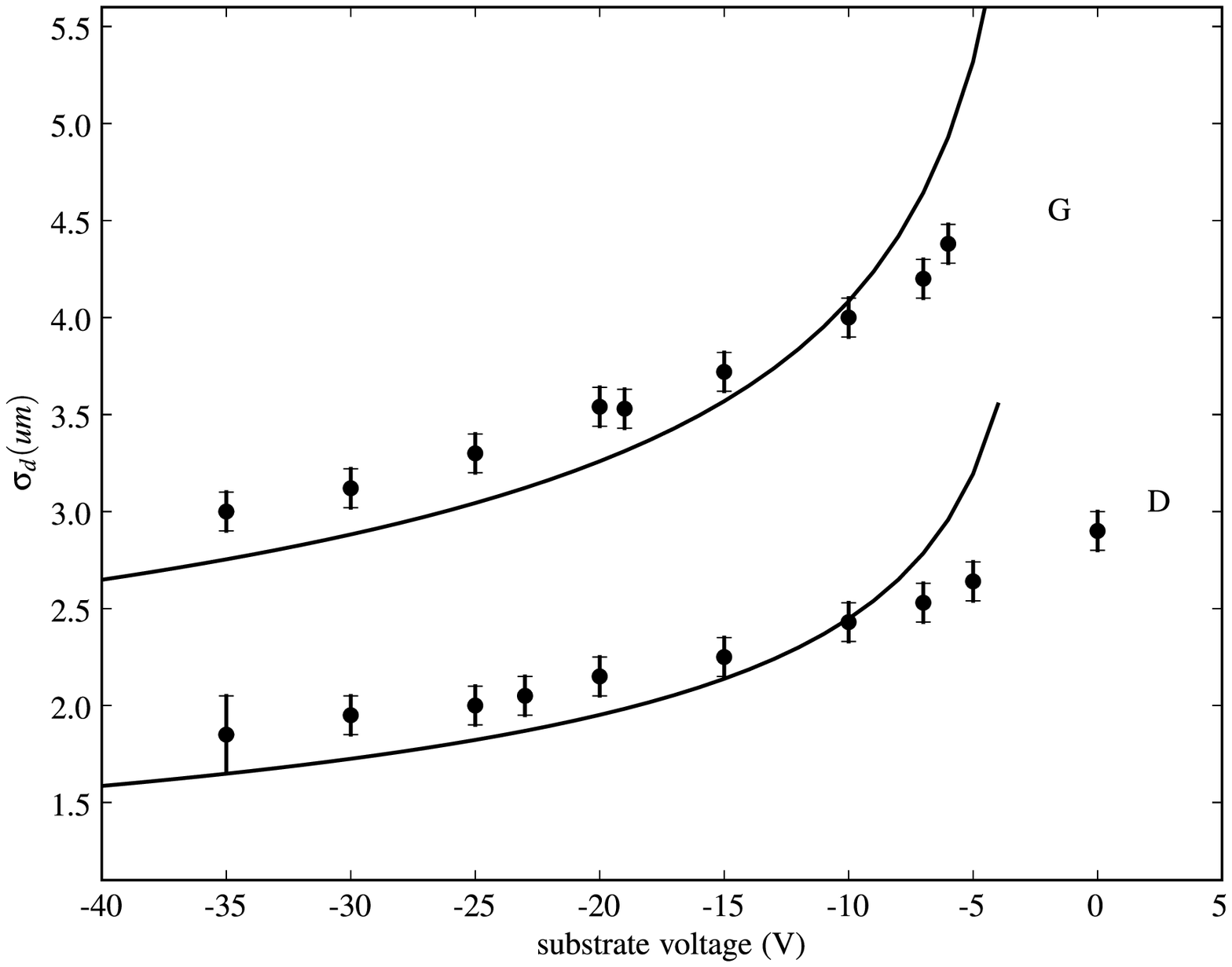}
    \caption{\small
      As in Figure \ref{fig:qsig_vsub}, but showing only $\sigma_{Q}$ for
      CCD-D and CCD-G.  The model of \citet{hop1987} (see Equations
      \ref{eq:hopmod} and \ref{eq:dd}) is overplotted for each set as a solid
      line.  For both devices we set  $V_{gate} = 8.5 V$ and we use
      a doping concentration of  $N_A = 8.9 \times 10^{11}$  for CCD-G
      and $N_A = 2.5 \times 10^{12}$ for CCD-D, based on the resistivities 
      shown in Table \ref{tab:ccdlist}. 
    \label{fig:hopmod}}
  \end{figure}

  \citet{hop1987} provides a theoretical model for the size 
  of a charge diffusion cloud, $\sigma$, as a function of the operating
  voltage, V. This model starts from the assumption 
  that the electron drift velocity is linearly proportional to the
  electric field strength at any height in the silicon $-$ an
  assumption we also adopt in Equation \ref{eq:sig_E}. 
  \citet{hop1987} predicts the final 1$\sigma$ radius of a charge cloud
  that begins at the surface of a CCD as: 
  \begin{equation}\label{eq:hopmod}
    \sigma_{d} = {\left[ {{2~kT~\epsilon}\over{e^2~N_A}}
      ~ln\left({ {d_d}\over{d_d - d}}\right) \right]}^{1/2} 
  \end{equation}
  \noindent where k is Boltzmann's constant, T is the temperature of
  the CCD, $\epsilon$ is the electric permittivity of silicon ($1.044
  \times 10^{-12}$ F cm$^{-1}$), e is the electron charge, and d is
  the thickness of silicon.
  $N_A$ is the concentration of acceptor atoms, which we determine
  from the resistivity values quoted in Table \ref{tab:ccdlist} using
  equation 1.24 from \citet[][p.75]{jan2001}. 
  $d_d$ is the depletion depth, given by:
  \begin{equation}\label{eq:dd}
    d_d = \left( {\frac{2V\epsilon}{e~N_A} } \right)^{1/2}
  \end{equation}
  where  V is the operating voltage of the device (approximately the
  gate voltage minus the substrate voltage). 
  Figure \ref{fig:hopmod} shows the data for CCD-D and 
  CCD-G with the above model overplotted as solid lines.  This figure
  demonstrates that the observed variation with $V_{sub}$ and d
  is reasonably well matched by the Hopkinson model for large values
  of the substrate voltage.  When $V_{sub}$ is small, the devices
  approach underdepletion, so that the depletion depth, $d_d$, is very
  close to the device thickness, d.  Hopkinson's model does not apply
  in this regime, which is why the data begin to deviate from the
  theory.

  \begin{figure}[tb]
    \centering
    \includegraphics[width=0.4\textwidth,draft=false]{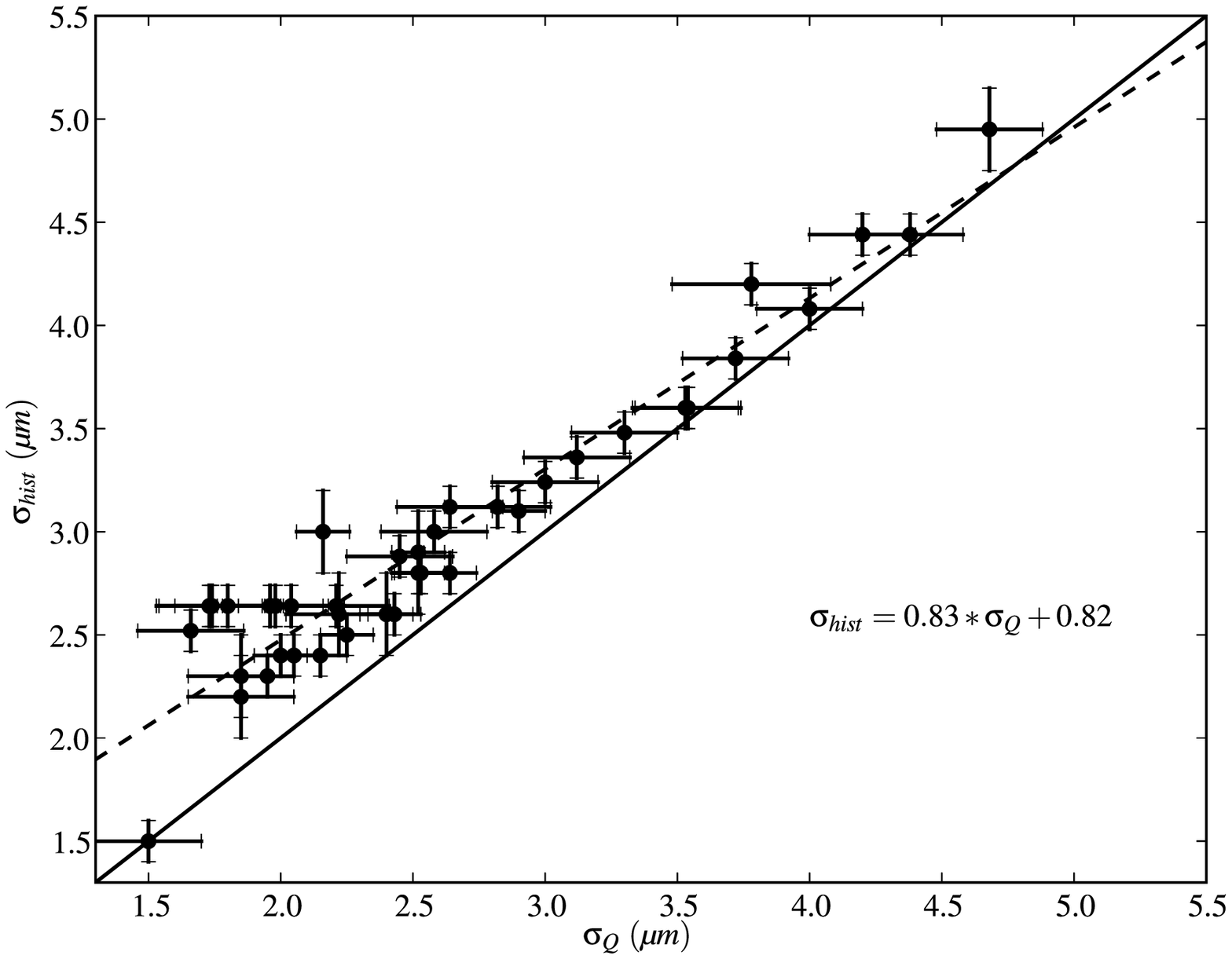}
    \caption{\small
      $\sigma_{Q}$ plotted against $\sigma_{hist}$ for all seven devices.
      The solid line shows where the points would fall if the two 
      methods were in perfect agreement. A least squares fit is
      plotted as the dashed line, with the functional form given in
      the lower right. 
    \label{fig:qsig}}
  \end{figure}

  When comparing the two methods against each other, we find that they
  produce consistent, if not identical results.  
  Figure \ref{fig:qsig} plots the measured value of $\sigma_{hist}$ 
  against $\sigma_Q$ for all seven devices. If the two methods measured
  precisely the same value for the physical diffusion length, the
  points would fall along the solid line in this figure with a slope
  of one.  A least squares fit to the points reveals a linear relation
  of the form $\sigma_{hist} = 0.83 \times \sigma_{Q} + 0.82$, shown
  as the dashed line.  As noted above, we have good reason to be
  suspicious of the accuracy of the histogram fitting model at very
  low values of $\sigma_d$. When $\sigma_d$ is less than
  about 0.25, the prominent histogram features become crowded, leading
  to an inaccurate estimate of the diffusion length. This deficiency
  could be corrected by employing substantially more x-rays in the
  histogram to improve energy resolution and signal to noise. 
  
  \section{Conclusions} \label{sec:conclusions}

  Using a set of four devices with a range of thicknesses and charge
  collection characteristics, we have evaluated two methods for using
  x-rays to measure the amount of charge diffusion in CCDs.  With a
  relatively simple model for the charge diffusion effect built off of
  the assumption of a Gaussian diffusion cloud, we are able to produce
  model  histograms that match the observed pixel energy distributions
  very accurately.  By varying the input parameters of the model to
  fit the observed data we can put precise constraints on both the
  amount of charge diffusion and the height of the barrier fields
  around each pixel.   We introduce the Q ratio to measure
  the amount of diffusion based on a simple quotient of the outer
  eight pixels to the central pixel in the summed x-ray PSF.  This
  method provides an even more direct means of assessing the diffusion
  properties of each device, as it requires no fitting procedure and
  is independent of model assumptions for relative diffusion
  measurements.  Using the same model as for the histogram fitting
  method, we are able to turn this simple Q ratio into an absolute
  measurement of the size of the diffusion cloud.  Comparing the
  fundamental diffusion parameters from each method, 
  $\sigma_{hist}$ and $\sigma_Q$,  shows that the two methods produce
  results that are in good agreement.   Due to the confusion of
  histogram features in low diffusion devices, 
  $\sigma_{hist}$ is best for measuring relatively large
  values of $\sigma_d$ and $\sigma_{Q}$ is more appropriate for
  the smaller values.  X-ray images are easy to
  collect in the lab, requiring 
  very little specialized equipment, and only minimal image processing
  is necessary to get the data needed for these techniques.  All of
  these advantages make the x-ray histogram and PSF ratio methods very
  useful "quick-look" additions to the CCD evaluation toolkit.

  \acknowledgements We would like to thank Sharon Erickson for
  invaluable assistance with laboratory equipment preparation, Gerry
  Luppino and Barry Burke for their help in procuring suitable test
  devices, and the referee for useful and constructive comments.
  

\clearpage

\clearpage

\clearpage

\clearpage

\clearpage

\clearpage

\clearpage

\clearpage

\clearpage

\end{document}